\documentclass[aps,prl,showpacs,twocolumn]{revtex4}

\usepackage{amsmath}
\usepackage{graphicx}
\usepackage{color}
\usepackage{subfigure}

\begin{document}

\title{Island size distributions in submonolayer growth: successful
  prediction by mean field theory with coverage dependent capture
  numbers}

\author{Martin K\"orner$^{1,2}$}
\author{Mario Einax$^{1}$}
\email{mario.einax@tu-ilmenau.de}
\author{Philipp Maass$^{2}$}
\email{philipp.maass@uni-osnabrueck.de}
\homepage{http://www.statphys.uni-osnabrueck.de}
\affiliation{$^{1}$Institut f\"ur Physik, Technische
Universit\"at Ilmenau,98684 Ilmenau, Germany\\
$^{2}$Fachbereich Physik, Universit\"at Osnabr\"uck,
Barbarastra{\ss}e 7, 49076 Osnabr\"uck, Germany}

\date{\today}

\begin{abstract}
  We show that mean-field rate equations for submonolayer growth can
  successfully predict island size distributions in the
  pre-coalescence regime if the full dependence of capture numbers on
  both the island size and the coverage is taken into account. This is
  demonstrated by extensive Kinetic Monte Carlo simulations for a
  growth kinetics with hit and stick aggregation. A detailed analysis
  of the capture numbers reveals a nonlinear dependence on the island
  size for small islands. This nonlinearity turns out to be crucial
  for the successful prediction of the island size distribution and
  renders an analytical treatment based on a continuum limit of the
  mean-field rate equations difficult.
\end{abstract}

\pacs{81.15.Aa,68.55.A-,68.55.-a}

\maketitle

The kinetics of submonolayer nucleation and island growth during the
initial stage of epitaxial thin film growth has been studied
intensively both experimentally and theoretically (for reviews, see
\cite{Evans/etal:2006,Michely/Krug:2004,Brune:1998}). A good
understanding of this kinetics assists in tailoring self-organized
nanostructures and thin film devices for specific needs. Mean-field
rate equations (MFRE) \cite{Venables:1973} successfully predict
important features such as the scaling behavior of the density of
stable islands with respect to the $\Gamma=D/F$ ratio of the adatom
diffusion rate $D$ and incoming flux $F$
\cite{Kandel:1997,Mueller/etal:1996,Einax/etal:2007,Einax/etal:2009}.
They seem to fail, however, to predict correctly the number densities
$n_s$ of islands composed of $s$ atoms, i.e.\ the island size
distributions (ISD) \cite{Bales/Chrzan:1994}. In this connection,
Ratsch and Venables \cite{Ratsch/Venables:2003} as well as Evans {\it
et al.} \cite{Evans/etal:2006} addressed a still open question:
whether the MFRE are successful in describing the precise shape of the
ISD, if the correct dependence of the capture numbers
$\sigma_s(\Theta)$ on both $s$ and the coverage $\Theta$ were taken
into account. The answer to this question is not obvious, since the
MFRE with correct capture numbers $\sigma_s(\Theta)$ still neglect (i)
many-particle correlation effects \cite{comm:corr}, (ii) spatial
fluctuations in shapes and capture zones of islands, and (iii)
coalescence events that, despite rare in the early-stage growth, could
have a significant influence.

Various theoretical approaches have been developed in the past for
obtaining appropriate analytical formulae or approximate numerical
results for the $\sigma_s(\Theta)$ (for details, see
\cite{Ratsch/Venables:2003,Evans/etal:2006} and references therein).
These approaches focus on the low-temperature case with critical size
$i=1$, i.e.\ the case when already dimers can be considered as stable
(on a time scale, where the ISD in the initial growth regime is
formed). The roughest approach is to neglect the $\Theta$ dependence
and to use just two numbers, $\sigma_1$ for the adatoms and an average
number $\bar\sigma$ for all stable islands with $s\ge2$, and to fit
these numbers to give best agreement with simulated or measured data.
Alternatively, simulated capture numbers for various $s$ at a fixed
coverage $\Theta$ have been considered \cite{Bartelt/Evans:1996} and
used in the analysis of experiments \cite{Bartelt/etal:1998}. As shown
in Fig.~1, however, neither of these approaches as well as a more
sophisticated self-consistent treatment
\cite{Bales/Chrzan:1994,Amar/etal:2001} is successful in providing a
good description of the ISD as obtained from KMC simulations. A first
numerical study for computing coverage dependent capture numbers has
been performed in \cite{Gibou/etal:2000,Gibou/etal:2003} using a level
set method. Integration of the MFRE with the obtained capture numbers
gave quantitative agreement with KMC results for the island density
$N$, but the statistics was insufficient to achieve conclusive answers
with respect to the ISD. For taking into account the correlation
between $s$ and the size of capture zone areas, i.e.\ that larger
islands tend to exhibit larger capture zones, a generalization of the
MFRE towards an evolution equation for the joint probability of island
size and capture area was set up
\cite{Mulheran/etal:2000,Amar/etal:2001,Evans/etal:2001}. This,
however, had to be done at the expense of introducing additional
parameters for considering nucleation events inside the capture zones.

\begin{figure}[b!] 
\centering
 \includegraphics[width=0.42\textwidth]{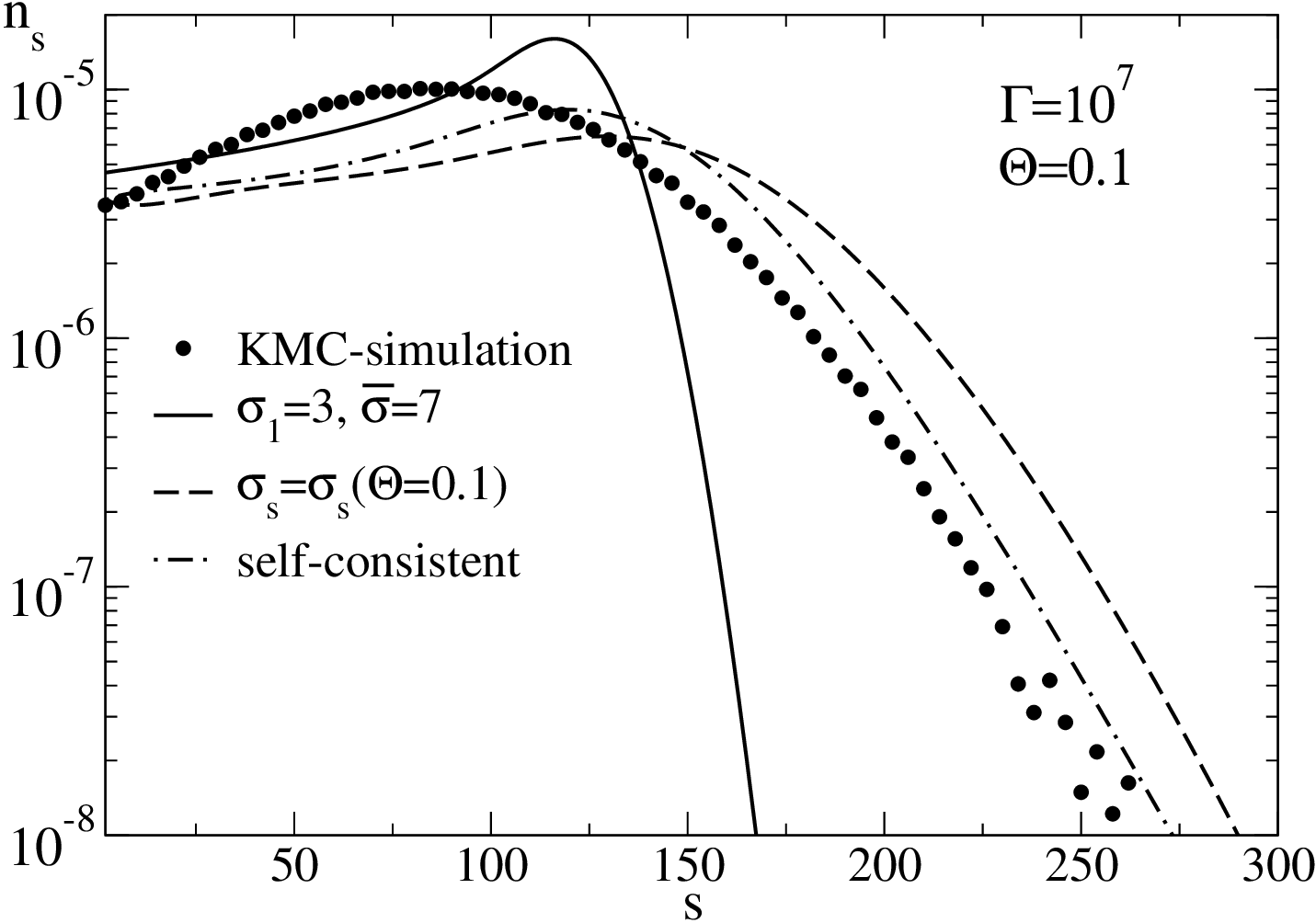}
 \caption{Island size distribution obtained from KMC simulation in
   comparison with ISDs calculated from an integration of the MFRE
   using three different approximations for the $\sigma_s(\Theta)$.}
\label{fig:fig1}
\end{figure}

In this Letter we compute the capture numbers $\sigma_s (\Theta)$ as a
function of both the island size $s$ and the coverage $\Theta$ by
performing extensive KMC simulations. We show that based on these
functions the ISD for growth kinetics with hit and stick aggregation is
well predicted by the MFRE in the growth regime before coalescence. We
discuss simplified forms of the capture numbers $\sigma_s(\Theta)$
with respect to predicting the ISD, which could render an analytical
treatment of the problem possible.

The MFRE for a situation at low temperatures (no re-evaporation) with
a critical nucleus of size $i=1$ and consideration of direct
impingement of arriving atoms at islands are
\begin{align}
\label{eq:n1}
\frac{d n_1}{d t} =& (1-\Theta) F - 2 D \sigma_1
n_1^2 - D n_1 \sum_{s>1} \sigma_s n_s\nonumber\\
&{}- 2 F \kappa_1 n_1 - F \sum_{s>1} \kappa_s
n_s \, \\
\frac{d n_s}{d t} =& D n_1 \left( \sigma_{s-1} n_{s-1} - \sigma_s
n_s \right)\nonumber\\
&{}+ F \kappa_{s-1} n_{s-1} - F \kappa_s n_s \,,\quad s=2,3,\ldots
\label{eq:ns}
\end{align}
These equations refer to the growth regime, where coalescence events
of islands should be negligible, and it is presumed that only single
adatoms are mobile and that atom movements between the first and
second layer can be disregarded. Moreover, adatoms arriving on top of
an island are not counted, i.e.\ $s$ in a strict sense refers to the
number of substrate sites covered by an island (or the island area).
Accordingly, the deposition flux $F$ of adatoms in Eq.~(\ref{eq:n1})
has to be restricted to the uncovered fraction $(1-\Theta)$ of the
substrate area. The terms $2D \sigma_1 n_1^2$ and $F\kappa_1n_1$
describe the nucleation of dimers due to attachment of two adatoms by
diffusion and due to direct impingement, respectively. The term $D
n_1\sigma_s n_s$ describes the attachment of adatoms to islands of
size $s>1$, and $F \kappa_s n_s$ the direct impingement of deposited
atoms to boundaries of islands with size $s$. Dividing
Eqs.~(\ref{eq:n1}) and (\ref{eq:ns}) by $F$ leads to evolution
equations with the coverage $\Theta=Ft$ as independent variable and to
a replace of $D$ by $\Gamma=D/F$ on the right hand side.

Our KMC simulations are performed with an exact continuous-time
algorithm and periodic boundary conditions for ``hit and stick''
aggregation on a square lattice with $L\times L=8000\times8000$ sites.
The lattice constant is set to unity. To calculate the capture numbers
$\sigma_s$ at the coverage $\Theta$, we use the following procedure:
Each simulation run is stopped at coverage $\Theta$ and the number
densities $n_s=N_s/L^2$, $s=1,2,\ldots$ are determined, where $N_s$
are the numbers of monomers ($s=1$) and islands ($s>1$). Then the
simulation is continued for a long time interval $T$ without
deposition and the following additional rules: (i) when an adatom is
attaching to an island of size $s>1$, a counter $M_s$ for such
attachments is incremented and the adatom thereafter repositioned at a
randomly selected site of the free substrate area (i.e.\ a site which
is neither covered nor a nearest neighbor of a covered site); (ii)
when two adatoms form a dimer, a counter $M_1$ for these nucleation
events is incremented and the two adatoms thereafter repositioned
randomly as described in (i). In this way a stationary state is
maintained at the coverage $\Theta$. Using the counters, the mean
times $\tau_s=T/M_s$, $s=1,2,\ldots$, for the respective nucleation
and attachment events are determined. Given these times, the capture
numbers $\sigma_s$ are calculated by equating $D\sigma_s n_1n_s$,
$s=1,2,\ldots$ with $1/\tau_s$, yielding $\sigma_s=1/[Dn_1n_s\tau_s]$.
Averaging the $\sigma_s$ over many simulation runs (configurations)
finally gives $\sigma_s(\Theta)$. The $\kappa_s(\Theta)$ are
determined from the lengths of the islands boundaries, which are
simultaneously monitored during the simulation and averaged for each
size $s$.

\begin{figure}[t!]
\centering
 \includegraphics[width=0.4\textwidth]{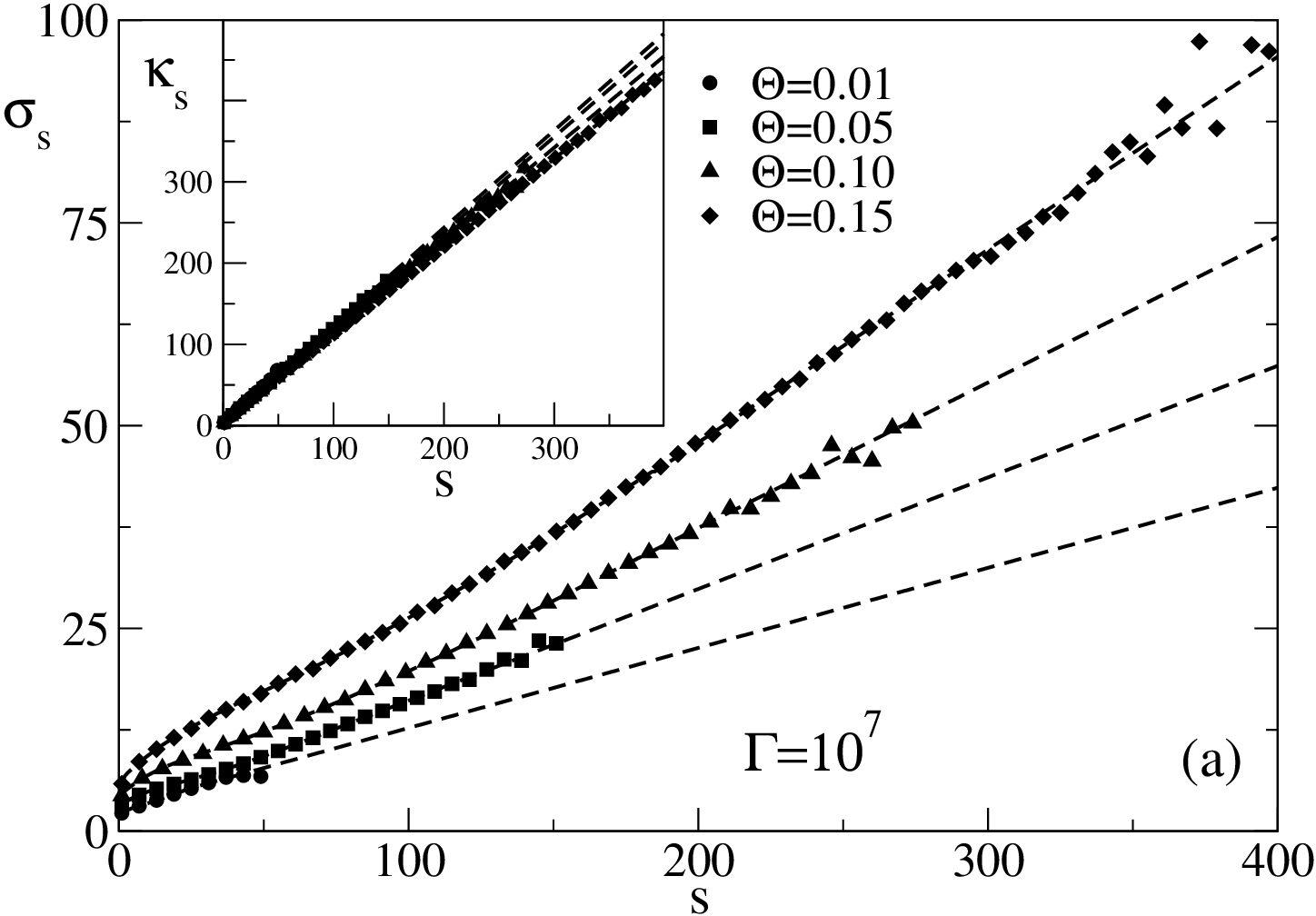}
 \hspace*{2ex}
 \includegraphics[width=0.41\textwidth]{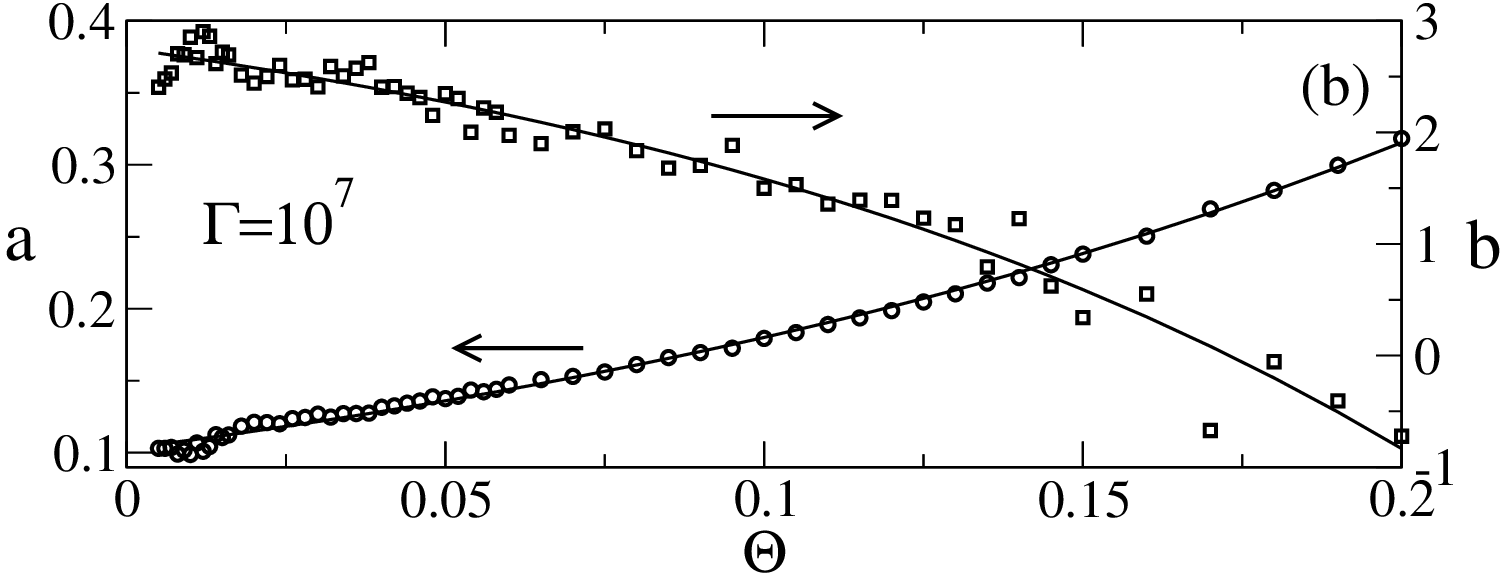}
 \hspace*{2ex}
 \includegraphics[width=0.41\textwidth]{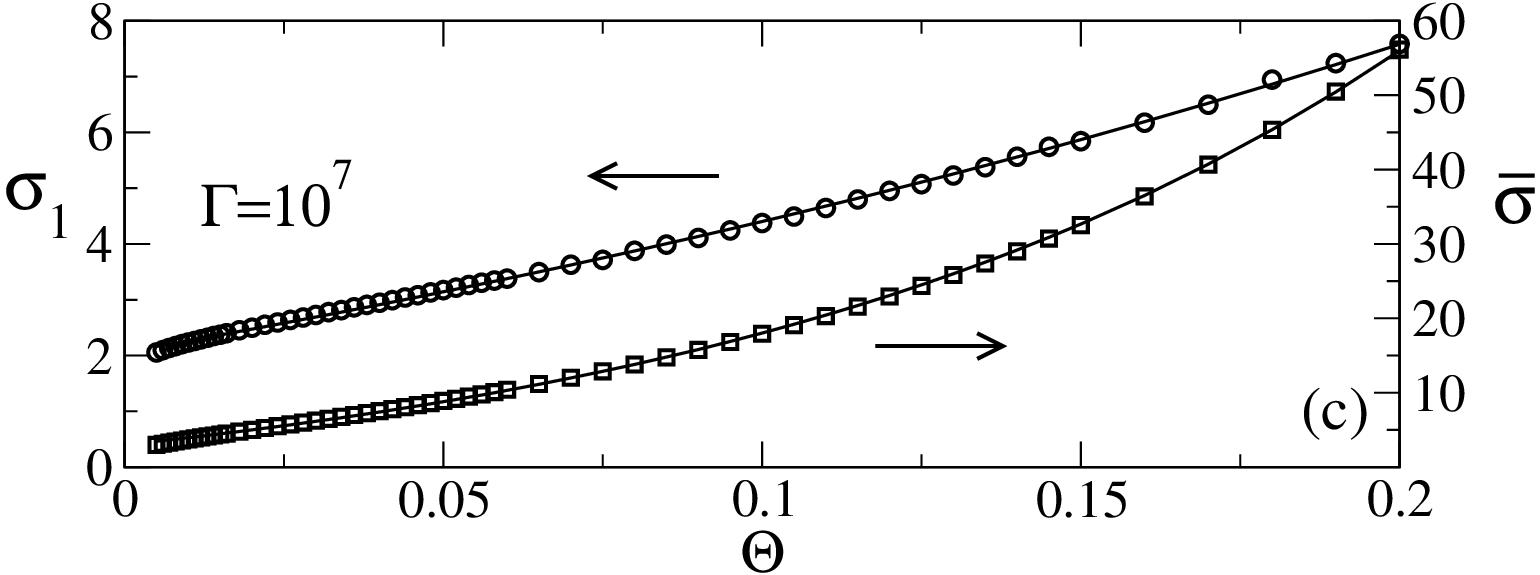}
 \caption{(a) Dependence of the capture numbers $\sigma_s(\Theta)$ on
   $s$ for four different fixed coverages; the inset shows the
   corresponding $\kappa_s(\Theta)$. (b) The coefficients $a$ and $b$
   of the asymptote $\sigma_s(\Theta)\sim a(\Theta)s+b(\Theta)$, and
   (c) $\sigma_1$ and $\bar\sigma$ as functions of $\Theta$. For a
   convenient extraction of the data in (b) and (c) the following fit
   function can be used (solid lines):
   $a=0.103\exp(5.6\Theta)$,
   $b=3.85-1.1\exp(7.26\Theta)$,
   $\sigma_1 = -4.5 + 6.55\exp(3.05\Theta)$, and
   $\bar\sigma(\Theta)=-6.8+9.8\exp(9.3\Theta)$.}
 \label{fig:fig2}
\end{figure}

Overall the functions $\sigma_s(\Theta)$ and $\kappa_s(\Theta)$ were
obtained for $57$ different $\Theta$ values in the range 0.005--0.2
and a large number of island sizes for each value of $\Theta$, ranging
up to $1000$ values for the largest $\Theta$. The typical number of
nucleation/attachment events for each $\Theta$ value was $10^8$.

Figure~\ref{fig:fig2}a) shows results for $\sigma_s(\Theta)$ as a
function of $s$ for four different fixed $\Theta$ at $\Gamma=10^7$.
For large $s$ we find a linear dependence of $\sigma_s(\Theta)$ on $s$
at all coverages, which can be explained \cite{Evans/etal:2006} by
noting that the $\sigma_s(\Theta)$ become proportional to the mean
capture zone areas $A_s$. Since a double-sized capture zone gives on
average rise to a double-sized island, it holds $A_s\sim s$ and hence
$\sigma_s\sim s$. The asymptotic behavior can be described by
$\sigma_s(\Theta)\sim a(\Theta)s+b(\Theta)$, where the slope
$a(\Theta)$ is an increasing and the offset $b(\Theta)$ a decreasing
function of $\Theta$, see Fig.~\ref{fig:fig2}b). For small $s$, a
nonlinear dependence of $\sigma_s(\Theta)$ on $s$ is found. As shown
in the inset of Fig.~\ref{fig:fig2}a), the direct capture numbers
$\kappa_s (\Theta)$ have also a linear dependence on $s$ but are
approximately independent of $\Theta$, i.e.\ $\kappa_s(\Theta)\simeq
s$. In Fig.~\ref{fig:fig2}c) we show the capture number $\sigma_1
(\Theta)$ related to nucleation events and the mean capture number
$\bar\sigma(\Theta) =\sum^{\infty}_{s=2}\sigma_s(\Theta) n_s/N$, where
$N=\sum_{s=2}^{\infty} n_s$. These functions are important when
considering the scaled capture numbers
$\sigma_s(\Theta)/\bar\sigma(\Theta)$ as function of the scaled island
size $s/\bar s(\Theta)$, where $\bar s(\Theta)= \sum^{\infty}_{s=2}s
n_s/N \cong 4.7+ 818\,\Theta$ at $\Gamma = 10^7$ here
\cite{comm:theta}).

\begin{figure}[t!]
\centering
 \includegraphics[width=0.4\textwidth]{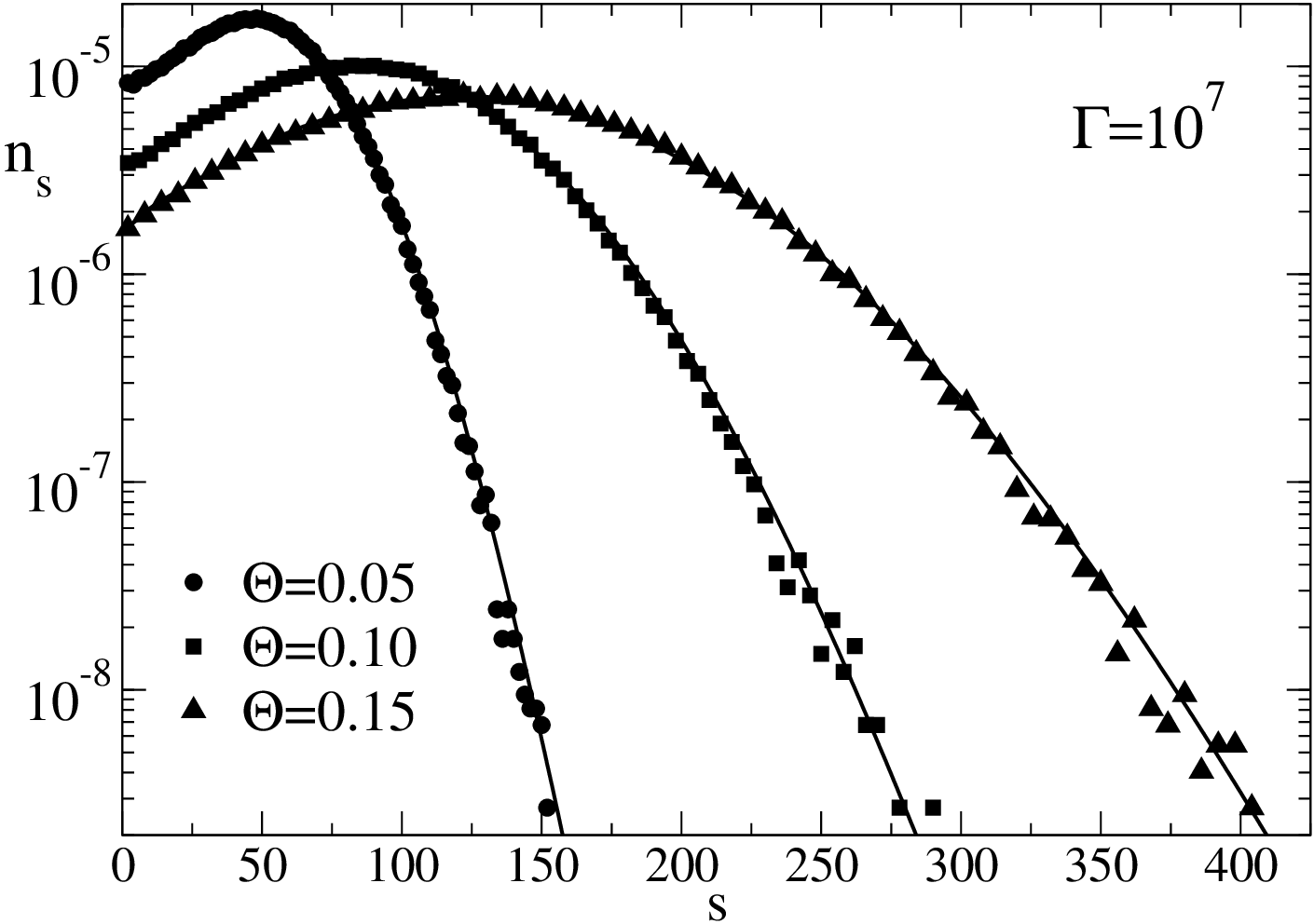}
 \hspace*{2ex}
 \includegraphics[width=0.4\textwidth]{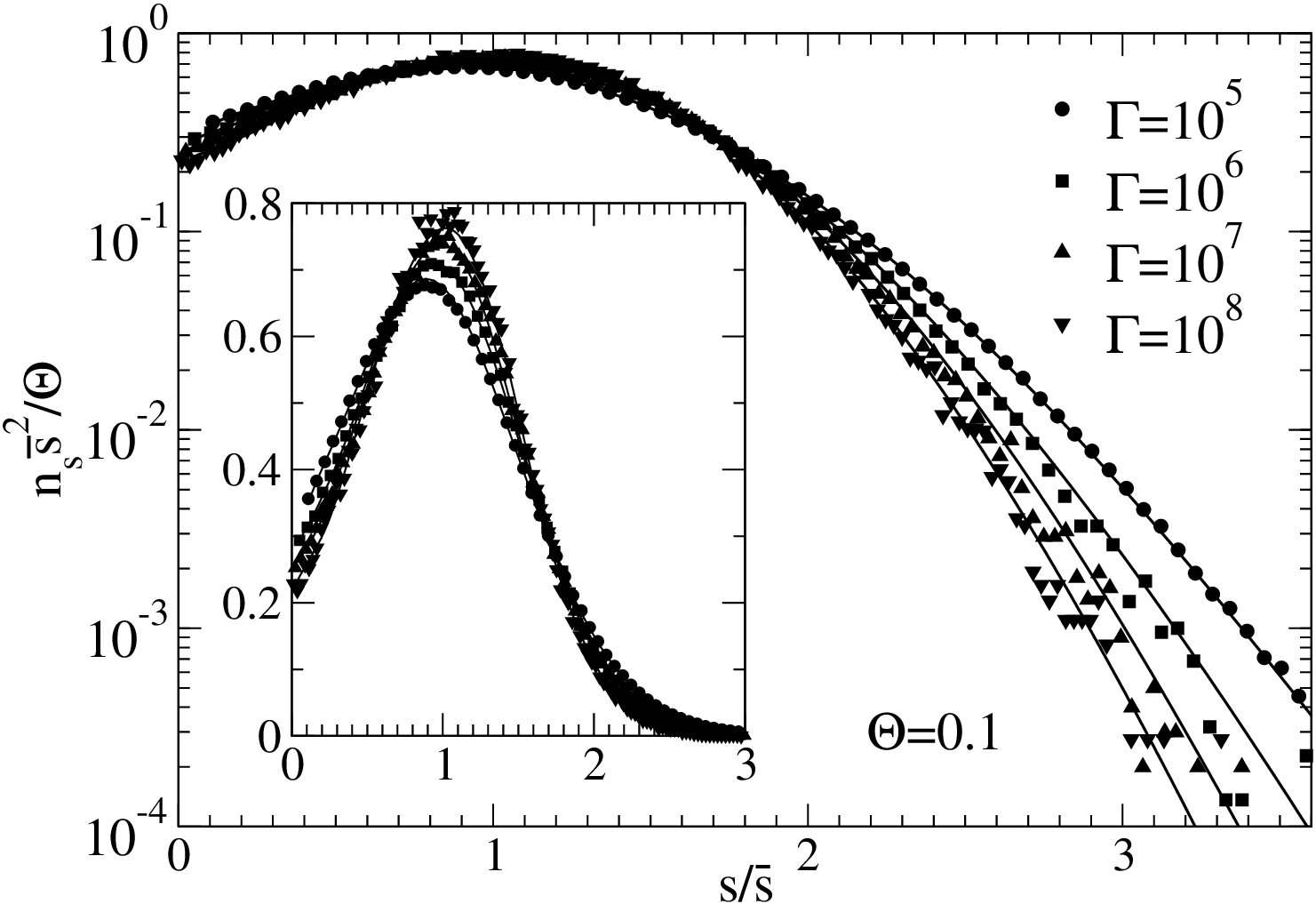}
 \hspace*{2ex}
 \caption{(a) Island size distributions for three different coverages
   at fixed $\Gamma=10^7$, and (b) scaled ISDs for four different
   $\Gamma$. The inset in (b) shows the scaled ISDs in a double-linear
   representation. The symbols mark the results from the KMC
   simulations and the solid lines the results obtained from
   integrating the MFRE equations (\ref{eq:n1}) and (\ref{eq:ns}) with
   $\sigma_s(\Theta)$ and $\kappa_s(\Theta)$ determined by KMC
   simulations (see text).}
 \label{fig:fig3}
\end{figure}

\begin{figure}[t!]
\centering
\includegraphics[width=0.4\textwidth,clip]{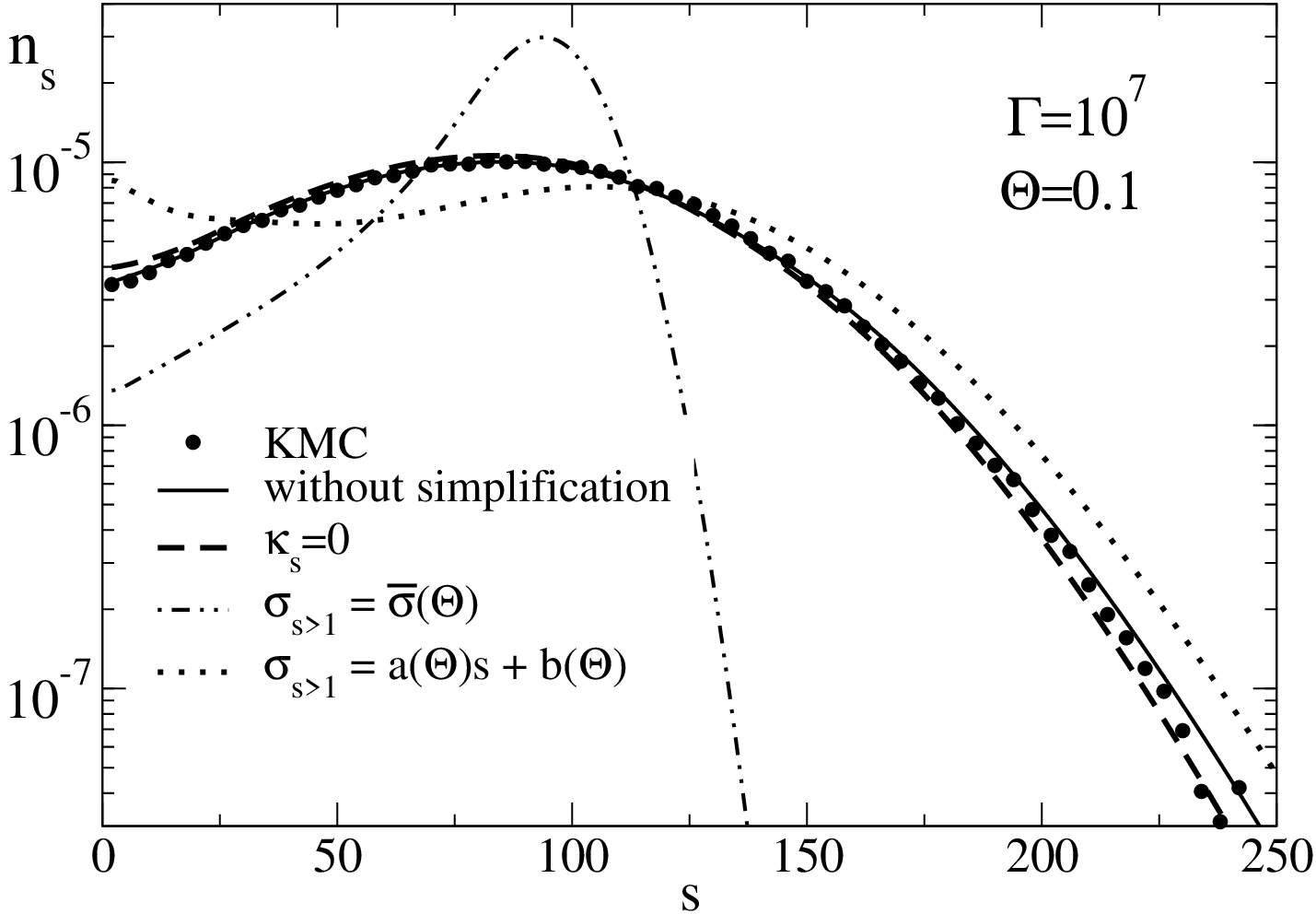}
\caption{Island size distribution for $\Theta=0.1$ from the KMC
  simulation in comparison with the MFRE results when using different
  simplifications with respect to the functional form of the capture
  numbers (see text).}
 \label{fig:fig4}
\end{figure}

By combining the linear function for large $s$ with a polynomial at
small $s$ to take into account the nonlinearity, we fitted the curves
in Fig.~\ref{fig:fig2}a) and used these fits to integrate the MFRE
(\ref{eq:n1}) and (\ref{eq:ns}). The data for the resulting MFRE-ISDs
in Figs.~\ref{fig:fig3}a,b are one of our key findings. As shown in
Fig.~\ref{fig:fig3}a, the MFRE-ISD (solid lines) is for all coverages
in the precoalescence regime in excellent agreement with the
corresponding KMC-ISD (symbols) obtained from the KMC simulations. A
variation of $\Gamma$ does not affect the quality of agreement, as can
be seen from Fig.~\ref{fig:fig3}b, where we plot the scaled ISDs
$n_s\bar s^2/\Theta$ versus $s/\bar s$ for a fixed coverage
$\Theta=0.1$ and four different $\Gamma$. Moreover, one can infer from
this figure that the scaled ISDs tends to approach a limiting master
curve when $\Gamma\to\infty$. For comparison with earlier results in
the literature, we show in the inset of Fig.~\ref{fig:fig3}b the
scaled ISDs in the more common double-linear plot instead of the
linear-log representation used otherwise in Figs.~\ref{fig:fig1},
\ref{fig:fig3}a,b, and \ref{fig:fig4}. We chose this linear-log
representation to show that the MFRE capture the behavior also
correctly in the wings at very small ($s\ll\bar s$) and very large
island sizes ($s\gg\bar s$). In fact, the agreement is seen over about
four orders of magnitude of $n_s$ in Fig.~\ref{fig:fig3}a. This
demonstrates that the approximations involved in the MFRE are
appropriate to predict the ISD with high accuracy for the hit and
stick aggregation considered here.

So far we have used the complete functional form for $\sigma_s
(\Theta)$ and $\kappa_s (\Theta)$. The question arises whether all
details seen in Fig.~\ref{fig:fig2}a) are necessary with respect to a
good prediction of the ISD. To this end we discuss the following
simplifications: {(i)} all $\kappa_s$ are set to zero, {(ii)} the
$\sigma_s (\Theta)$ are replaced by $\sigma_1(\Theta)$ for $s=1$ and
$\bar\sigma(\Theta)$ for $s\geq2$ (and analogously for the $\kappa_s
(\Theta)$), and {(iii)} the asymptotics $\sigma_s(\Theta)\sim
a(\Theta)s+b(\Theta)$ is used for all $s\geq2$, while we keep the
$\sigma_1(\Theta)$ (again the analogous procedure is used for the
$\kappa_s(\Theta)$).

Figure~\ref{fig:fig4} shows the MFRE-ISD resulting from these
simplifications. Neglecting the $\kappa_s$ in Eqs.~(\ref{eq:n1}) and
(\ref{eq:ns}), the ISD is again well predicted, see the dashed line.
For increasing $\Gamma$ the agreement becomes even better (not shown).
When neglecting the $s$-dependence (case (ii)) the MFRE-ISD has a
maximum still close to the KMC-ISD, but its width is much smaller than
that of the KMC-ISD. The width of the the respective scaled
distribution tends to zero for $\Gamma\to\infty$. Let us remind that
we already showed in Fig.~\ref{fig:fig1}) that a full neglect of the
$\Theta$ dependence also does not yield a good ISD. In case (iii) the
MFRE-ISD is also poor in comparison with the KMC-ISD. The MFRE-ISD
shows a second maximum at $s=2$, which is caused by the fact that the
linear relationship underestimates the $\sigma_2(\Theta)$ value,
leading to a higher lifetime and correspondingly larger concentration
of dimers. Generally speaking, a linear relationship between
$\sigma_s(\Theta)$ and $s$ does not cover the small $s$ behavior but,
as one would expect, it gives a fair account of the shape of the ISD
for large $s$.

In summary, we have demonstrated that an integration of the standard
MFRE with coverage-dependent capture numbers yields an MFRE-ISD that
for hit-and-stick aggregation is in excellent agreement with the
KMC-ISD. The full dependence of the capture numbers on both the island
size and the coverage was determined from extensive KMC simulations
and the functional form was analyzed in detail. Despite the fact that
a linear dependence on the island size holds over almost the entire
$s$-range, the nonlinear behavior is crucial for a good account of the
ISD. This implies that it will be difficult to find simple functions,
which one could use in an analytical continuum approach for the scaled
ISD \cite{Bartelt/Evans:1996}.

We thank H.~Brune, W.~Dieterich, and J.~Evans for very valuable
discussions.


\begin{thebibliography}{99}
\bibitem{comm:corr} When starting from a many-particle master equation
  for the kinetics, nonlinear terms $\propto n_1n_s$, cf.\
  Eq.~(\ref{eq:ns}), would arise from factorizing non-equlibrium
  correlators of occupation numbers.
\end{thebibliography}
\end{document}